\documentclass[12pt,fleqn]{article}

\usepackage{amsmath, amssymb,graphicx}
\newcommand \be{\begin{equation}}
\newcommand \ee{\end{equation}}
\newcommand \bea{\begin{eqnarray}}
\newcommand \eea{\end{eqnarray}}
\newcommand \bee{\begin{equation}}
\newcommand \eee{\end{equation}}

\newcommand\TL{\hfil$\displaystyle{##}$}
\newcommand\TR{$\displaystyle{{}##}$\hfil}
\newcommand\TC{\hfil$\displaystyle{##}$\hfil}

\def\seqalign#1#2{\vcenter{\openup1\jot
  \halign{\strut #1\cr #2 \cr}}}
\def\lbldef#1#2{\expandafter\gdef\csname #1\endcsname {#2}}
\newcommand{\eqn}[3][]{\lbldef{#2}{(\ref{#2})}
\def\@eqnstyle{#1}
\ifx\@eqnstyle\@empty
\begin{equation} \eqalign{#3} \label{#2} \end{equation}
\else
\begin{equation} \seqalign{\span\TC}{#3} \label{#2} \end{equation}
\fi}
\def\eqalign#1{\vcenter{\openup1\jot
    \halign{\strut\span\TL & \span\TR\cr #1 \cr
   }}}
\def\eno#1{(\ref{#1})}
\def\mop#1{\mathop{\rm #1}\nolimits}
\def\tr{\mop{tr}}

\def\be{ \begin{equation} }
\def\ee{ \end{equation} }

\input epsf

\begin{document}

\thispagestyle{empty}
\renewcommand{\thefootnote}{\fnsymbol{footnote}}

{\hfill \parbox{4cm}{ hep-th/0604058 \\
PUPT-2194\\Brown HET-1467\\ ITEP-TH-12-06}}

\bigskip\bigskip

\begin{center} \noindent \Large \bf
 Calibrated Surfaces  and \\ Supersymmetric Wilson Loops
\end{center}

\bigskip\bigskip\bigskip

\centerline{ \normalsize \bf A. Dymarsky,$^{a}$ S. Gubser,$^{a}$
Z. Guralnik,$^{b,c}$ and J. Maldacena$^{d}$\footnote[1]{\noindent
\tt dymarsky@princeton.edu, ssgubser@princeton.edu,
zack@het.brown.edu, malda@ias.edu} }

\bigskip
\bigskip\bigskip

\centerline{$^a$ \it Joseph Henry Laboratories, Princeton
University, Princeton, NJ 08544, USA}
\bigskip
\centerline{$^b$ \it Department of Earth Atmospheric and Planetary
Science,}\centerline{\it Massachusetts Institute of Technology,
Cambridge MA 02139, USA} \centerline{$^c$ \it Department of
Physics, Brown University, Providence RI, 02912, USA}
\bigskip
\centerline{$^d$ \it Institute for Advanced Study, Princeton NJ
08540, USA.}
\bigskip\bigskip

\bigskip\bigskip

\renewcommand{\thefootnote}{\arabic{footnote}}

\centerline{\bf \small Abstract}
\medskip

{\small We study the dual gravity description of
 supersymmetric Wilson loops whose expectation value is unity. They are described by calibrated surfaces that end on
 the boundary of anti de-Sitter space and are
pseudo-holomorphic with respect to an almost
complex structure on an eight-dimensional slice of $AdS_5 \times
S^5$.  The regularized area of these surfaces vanishes, in
agreement with field theory  non-renormalization theorems for the
corresponding operators.}

\newpage

\section{Introduction}
\label{INTRODUCTION}

At $N\rightarrow\infty$ and $\lambda =g^2 N \gg 1$, the
expectation value of a special class of Wilson loops in ${\cal
N}=4$ super Yang-Mills can be computed using AdS/CFT
duality \cite{Maldacena:1998im,Rey:1998ik,Drukker:1999zq}.  These
Wilson loops are sometimes referred to as ``locally BPS,'' and they are
naturally identified with a loop at the boundary of $AdS_5 \times
S^5$.  Their expectation value is computed by evaluating the
semiclassical partition function for a string with boundary on
this loop.  At leading order we have to find the minimal area of
the worldsheet with these boundary conditions.

There are conjectured all-order results for a subset of the
locally BPS Wilson loops.  These include certain circular Wilson
loops which are invariant under a combination of Poincar\'e and
conformal supersymmetries \cite{Drukker:2000rr} (see also
\cite{morecomplex}). The latter are related by an singular
conformal transformation (inversion) to an infinitely extended
straight Wilson line which is invariant under $1/2$ of the global
Poincar\'e supersymmetries. The latter is not renormalized,
meaning it has expectation value $1$.

In \cite{Zarembo:2002an} an interesting class of supersymmetric
Wilson loops was introduced.  These are loops where the direction
on the $S^5$ at each point along the loops varies according to the
direction in ${\bf R}^4$ of the tangent vector ${\dot x^\mu \over |\dot
x^\mu| }$.  There it was also conjectured that the Wilson loops
which are invariant with respect to $1/4$ of the Poincar\'e
supersymmetries are also not renormalized, at least in the large
$N$ limit.  This proposal was based on a comparison of $\lambda
\ll 1$ perturbative results with $\lambda \gg 1$ results obtained
using AdS/CFT duality.  The latter were obtained for the specific
case of circular and infinitely extended rectangular Wilson loops,
in which case the minimal surfaces are known.  Using field
theoretic arguments, this non-renormalization theorem was proven
in \cite{GK,Guralnik:2004yc} and extended to the case of $1/8$
supersymmetric Wilson loops as well as finite $N$.  We will extend
these arguments for all SUSY loops by arguing that the Wilson
loops are BRST trivial operators in a topologically twisted
theory.

The non-renormalization of  supersymmetric Wilson loops implies
that the associated minimal surfaces in $AdS_5 \times S^5$ have
zero regularized area. More precisely, the area of the minimal
surface must be equal to a divergent term proportional to the
perimeter of the Wilson loop, with no additional finite parts.
This has only been explicitly shown for the circular and
infinitely extended rectangular loops. The purpose of this article
is to study the more general case.

We will demonstrate the existence of a calibration two-form
$J_{mn}dx^m\wedge dx^n$, for which the associated minimal surfaces
have boundary behavior corresponding to supersymmetric Wilson
loops.  Furthermore, the area of these surfaces, $A= \int J $, is
exactly equal to the divergent term proportional to the perimeter
of the Wilson loop.  These surfaces are pseudo-holomorphic curves
with respect to an almost complex structure given by $J^m{}_n$.
Complex surfaces in $AdS$ were also studied in connection with
baryons and various other objects, see \cite{uranga} and
references therein.

The organization of this article is as follows.  In
section~\ref{WILSON}, we review some basic features of BPS Wilson
loops.  In section~\ref{MINIMAL}, we review the computation of
Wilson loop expectation values in ${\cal N}=4$ SYM using minimal
surfaces in $AdS_5 \times S^5$.  In section~\ref{CALIBRATE}, we
find a calibration two-form for which the associated minimal
surfaces are pseudo-holomorphic curves and have boundary behavior
corresponding to supersymmetric Wilson loops.  In
section~\ref{COUNTING}, we discuss the existence and multiplicity
of solutions for fixed boundary conditions.  In section~\ref{SUSY}
we show the surfaces preserve some supersymmetries.  In
section~\ref{ANNULUS} we consider a worldsheet solution which
arises when we have coincident circular Wilson loops, as well as certain
other solutions with a $U(1)$ isometry.

\section{BPS Wilson loops}
\label{WILSON}

It is natural to consider Wilson loops in ${\mathcal N} =4$ gauge
theory which involve the adjoint scalars $\phi^a$ as well as the
gauge fields.  The Wilson loop which is usually considered has
(in Euclidean space) the form
\begin{align}\label{eucloop}
W = \frac{1}{N}\tr  P \exp  i \int ds \left( A_{\mu}
\frac{dx^{\mu}(s)}{ds} + i\phi^a \frac{d\zeta^a (s)}{ds} \right) \,,
\end{align}
where $a =1 \cdots 6$ and $\zeta^a(s)$ is a path in an auxiliary
space which, unlike $x^\mu(s)$, is not necessarily closed. A
special class of loops satisfying $|\dot{\zeta}| = |\dot{x}|$ can
be identified with loops at the boundary of $AdS_5 \times S^5$.
The latter can be written as
\begin{align}\label{lbpsloop}
W = \frac{1}{N}\tr  P \exp  i \int ds \left( A_{\mu}
\frac{dx^{\mu}(s)}{ds} + i\phi^a  \hat \theta^a (s) |\dot x|
\right) \,,
\end{align}
where $\hat\theta^a \hat\theta^a =1$, and the path at the boundary
of $AdS_5 \times S^5$ is given by $x^\mu(s), \hat\theta^a(s)$.
These Wilson loops are sometimes referred to as ``locally BPS''
because at each point $s$ along the loop they are invariant under
certain $s$-dependent supersymmetry transformations $\epsilon(s)$
satisfying
 \begin{align}\label{sus}
  \big[ \gamma^\mu \dot x^\mu(s) + i \gamma^a \hat\theta^a(s)|\dot x|
   \big]
   \epsilon(s) =0 \,,
 \end{align}
where $\gamma^\mu, \gamma^a$ are ten-dimensional gamma matrices.
Solutions to this equation exist because $\gamma^\mu \dot x^\mu(s)
+ i \gamma^a \hat\theta^a(s)|\dot x|$ is nilpotent. We will be
interested in Wilson loops that preserve a Poincar\'e supersymmetry.
This will happen only if there is a common solution $\epsilon(s)
= \epsilon$, independent of $s$.  Then the Poincar\'e supersymmetry associated with $\epsilon$ will be preserved.  A simple way to ensure this condition is to set
 \begin{align}\label{susy}
  \hat\theta^a(x) = \delta^a_\mu \dot x^\mu(s)/|\dot x| \,.
 \end{align}
Pairing values of $a$ with $\mu$ through $\delta^\mu_a$ as in (\ref{susy}) clearly involves some arbitrariness: an $SO(4) \times SO(6)$ transformation may be applied to obtain variants of (\ref{susy}) with identical properties.  Solutions to \eqref{sus} with constant $\epsilon(s) = \epsilon$ were enumerated in \cite{Zarembo:2002an}.  The number of unbroken supersymmetries depends on the dimension of the plane in which the path $x^\mu(s)$ lives.  A generic path lies in ${\bf R}^4$ and preserves $1/16$ of the Poincar\'e supersymmetries.  If the path $x^\mu(s)$ lies in an ${\bf R}^3$, ${\bf R}^2$, or ${\bf R}^1$ sub-plane, then $1/8$, $1/4$, or $1/2$ of the supersymmetries are unbroken, respectively.

The Wilson loop operators obeying  (\ref{susy}) also arise when one considers topologically twisted theories.  One of the topological twists of ${\cal  N} =4$ SYM consists in embedding the $SO(4)$ spin connection into an $SO(4)$ subgroup of $SO(6)$, namely the $SO(4)$ that rotates the first four transverse directions \cite{sofourtwist}.  Under this topological twist we have two spinors that are singlets of the twisted Lorentz group.  The generic BPS Wilson loop preserves only one of these spinors.  Under this twisting, four of the scalars can be naturally viewed as one forms.  Then the Wilson operator (\ref{lbpsloop})(\ref{susy}) arises when we consider the complex one forms $A_a + i \phi_a$.  We can similarly consider twisted, or partially twisted, theories where the spin connection along $n$ of the world-volume directions is embedded into the $SO(n)$ rotating $n$ of the transverse directions. Such twistings arise, for example, when we consider a special lagrangian $n$-cycle embedded in an $n$-complex dimensional Calabi Yau space \cite{vafatwist}.  The Wilson loop operators we are considering preserve the same supersymmetry; thus they are candidate operators of the topological theory.  More precisely, the BRST operator $Q$ of the topological theory annihilates these Wilson loop operators.  Formally these Wilson loops are also BRST trivial (see Appendix A) if the loop is is defined on a topologically trivial cycle.  In the case of loops on ${\bf R}^4$ all the cycles are, at least formally, topologically trivial.

The $1/8,1/4$ and $1/2$ BPS loops are special in the sense that
they can be written as bottom components of chiral superfields
with respect to a four supercharge subalgebra of the full ${\cal
N}=4$, $D=4$ supersymmetry \cite{GK}.  This subalgebra---with ${\cal N}=4, D=1$---involves supercharges whose commutators only give translation in one dimension.  The gauge connections in the remaining three directions belong to chiral superfields $\Phi_i$, with bottom
components $\Phi_i|_{\rm bot} = A_i + i\varphi_i$, $i=1,2,3$.  The
bottom component of the chiral Wilson loop,
 \begin{align}
  \frac{1}{N}\tr \, \left({\cal P}\, e^{i\int \Phi_i
    dx^i} \right) \,,
 \end{align}
is precisely of the form \eqref{lbpsloop} satisfying the
supersymmetry condition \eqref{susy}. The fact that these Wilson
loops can be written as bottom components of chiral superfields
was used in \cite{GK} to show that they have expectation value
$1$.  We will see below how this result is obtained in the dual AdS description.

\section{Minimal surfaces in $AdS_5 \times S^5$}
\label{MINIMAL}

The expectation value of locally BPS Wilson loops is computed in
AdS by evaluating the partition function for a string with
boundary conditions determined by $x^{\mu}(s)$, $\hat\theta^a(s)$.
The $AdS_5 \times S^5$ background is
 \begin{align}
  ds^2 &= \frac{R^2}{Z^2} \left( dX^\mu dX^\mu +
    \sum_{m=1}^6 dZ^m dZ^m \right) \nonumber \\
  Z^2 &\equiv Z^m Z^m,\qquad R^2= \sqrt{\lambda}\alpha' \,.
 \end{align}
The boundary of AdS is $Z=0$ but, for purposes of regularization, we will set it at $Z=\epsilon$.  The boundary conditions for the worldsheet associated to locally BPS Wilson loops are
 \begin{align}\label{BCS}
  X^{\mu}(\sigma =0,\tau) &= x^\mu(\tau)\nonumber\\
  \frac{1}{\epsilon} Z^m(\sigma =0, \tau)  &=
    \hat \theta^m(\tau)= \delta_\mu^m\frac{\dot x^\mu}{|\dot x|}
    \,.
 \end{align}
To leading order in the $1/\sqrt{\lambda}$ (or $\alpha'$) expansion,
the Wilson loop vacuum expectation value is given by the
semiclassical disc partition function:
\begin{align}\label{semicl} \langle W \rangle = {\cal V} e^{-{\cal A}_{\rm reg}} \,. \end{align}
The quantity ${\cal A}_{\rm reg}$ is found by minimizing a certain
Legendre transform of the area (see \cite{Drukker:1999zq}) which,
for the boundary conditions associated to locally BPS Wilson loops,
is equivalent to a regularization of the area of the minimal
surface. For these boundary conditions,  the minimal area has the
form
 \begin{align}
  A_{\rm min} = {\cal A}_{\rm reg} +
   \frac{1}{\epsilon} \oint d\tau|\dot x|\,,
 \end{align}
where ${\cal A}_{\rm reg}$ is finite as $\epsilon\rightarrow 0$. The
minimal surface is not necessarily unique and the pre-factor ${\cal
V}$ in \eqref{semicl} is a power of $1/{\sqrt{\lambda}}$ which
depends on the number of collective coordinates (see
\cite{Zarembo:2002an}).

For a non-renormalized Wilson loop we expect  $\langle W \rangle =
{\cal V} \exp(-{\cal A}_{\rm reg}) = 1$.  It has been shown
\cite{Zarembo:2002an} that ${\cal A}_{\rm reg} = 0$ for the minimal
surface associated to
 the circular supersymmetric
Wilson loop, with
 \begin{align}
  \eqalign{
  (\dot x^1, \dot x^2, \dot x^3, \dot x^4) &=
    (-\rho\sin\varphi, \rho\cos\varphi, 0, 0)  \cr
  (\hat\theta^1,\hat\theta^2,\hat\theta^3,\hat\theta^4,
    \hat\theta^5,\hat\theta^6) &=
   (-\sin\varphi , \cos\varphi, 0,0,0,0) \,, } \label{circular}
 \end{align}
as well as the infinitely extended rectangular Wilson loop. The
number of collective coordinates in these cases turns out to be
three \cite{Zarembo:2002an}, such that ${\cal V}=1$.
Non-renormalization in the rectangular case corresponds to the
no-force condition between BPS objects with the same charges. We
wish to compute the area of the minimal surface for more general
boundary shapes satisfying the supersymmetry conditions (\ref{BCS}).
For non-renormalized supersymmetric Wilson loops, one expects
 \begin{align}\label{div}
  A_{\rm min} = \frac{1}{\epsilon} \oint d\tau |\dot x| \,.
 \end{align}
Finding the minimal surfaces for a generic boundary
shape is a difficult problem.  However the problem
simplifies for calibrated surfaces.  We shall show below that a
calibration two-form $J$ exists for which smooth calibrated
surfaces satisfy the supersymmetric boundary conditions
(\ref{BCS}).  The area of these surfaces is precisely \eqref{div}.
Furthermore $J^m{}_n$ is an almost complex structure with
respect to which the calibrated surfaces are pseudo-holomorphic.

\section{Pseudo-holomorphic curves in $AdS_5 \times S^5$}
\label{CALIBRATE}

It is convenient to work with coordinates $\mathbb{X}^M = (X^\mu,Y^m, U^i)$ with $M=1\cdots10$,
$\mu =1\cdots4$, $m=1\cdots4$, $i=1,2$, in terms of which the $AdS_5 \times S^5$ geometry is
 \begin{align}\label{adsgeom}
  ds^2 = G_{MN} d \mathbb{X}^M \mathbb{X}^N =
   {Y^2+U^2 \over R^2} dX^\mu dX^\mu +
   \frac{R^2}{Y^2+U^2}(dY^m dY^m + dU^i dU^i).
 \end{align}
For the closed two form
 \begin{align} J =
  J_{AB}d\mathbb{X}^A\wedge d\mathbb{X}^B = \delta_{\mu m}dX^\mu
  \wedge dY^m \,,
 \end{align}
one finds that
 \begin{align} \label{jsq}
  J_A{}^B J_B{}^C = - \delta_A^\mu\delta^C_\mu -
  \delta_A^m\delta^C_m \,.
 \end{align}
Thus $J_A{}^B$ is an almost complex structure\footnote{The
Nijenhuis tensor does not vanish.  The surfaces of constant $U^i$
are almost-Kahler manifolds.} on surfaces of constant $U^i$.

There are two-dimensional minimal surfaces in $AdS_5 \times S^5$
which are pseudo-holomorphic with respect to the almost complex
structure $J_A{}^B$, and calibrated with respect to the two-form
$J$.  To see this, consider the positive quantity
 \begin{align}\label{ThisPos}
  {\cal P} &= {1 \over 4} \int d^2 \sigma \, \sqrt{g} g^{\alpha\beta}
    G_{MN} v_\alpha^M v_\beta^N  \cr
  v_\alpha^M &\equiv \partial_\alpha \mathbb{X}^M - \kappa J^M{}_N
   {\jmath}_\alpha{}^\beta \partial_\beta \mathbb{X}^N\end{align}
where $\kappa = \pm 1$, $g_{\alpha\beta}$ is an arbitrary positive definite metric on the worldsheet $\Sigma$ and
$\jmath_\alpha{}^\beta$ is the complex structure on $\Sigma$, namely
$\jmath^{\alpha\beta} = \epsilon^{\alpha\beta}/\sqrt{g}$ where $\epsilon^{12} = -\epsilon^{21} = 1$.  Expanding things out gives
 \begin{align}\label{ExpandOut}
  {\cal P} &= {1 \over 4} \int d^2 \sigma \, \sqrt{g} g^{\alpha\beta}
   G_{MN} \big( \partial_\alpha \mathbb{X}^M \partial_\beta \mathbb{X}^N +
    \kappa^2 J^M{}_K J^N{}_L \jmath_\alpha{}^\gamma
     \jmath_\beta{}^\delta \partial_\gamma \mathbb{X}^K \partial_\delta \mathbb{X}^L
      \cr &\qquad\quad{} -
     2\kappa J_{MN} \jmath^{\alpha \beta}  \partial_\alpha \mathbb{X}^M
      \partial_\beta \mathbb{X}^N \big)  \cr
   &= {1 \over 2} \int d^2 \sigma \, \sqrt{g} g^{\alpha\beta} G_{MN}
     \partial_\alpha \mathbb{X}^M \partial_\beta \mathbb{X}^N  -
    {\kappa \over 2} \int d^2 \sigma \, \sqrt{g} \jmath^{\alpha\beta}
     J_{MN} \partial_\alpha \mathbb{X}^M \partial_\beta \mathbb{X}^N \cr
     &\,\,\,\,\,\,\qquad - \frac{1}{4}\int d^2 \sigma
     \, \sqrt{g} g^{\alpha\beta}G_{ij}\partial_\alpha U^i \partial_\beta U^j \cr &=
     {\rm Area}(\Sigma) - \kappa\int_\Sigma J \,\,\,- \frac{1}{4}\int d^2 \sigma
     \, \sqrt{g} g^{\alpha\beta}\frac{1}{Y^2 + U^2}\partial_\alpha U^i \partial_\beta U^i \,.
 \end{align}
We have used $g^{\alpha\beta} \jmath_\alpha{}^\gamma \jmath_\beta{}^\delta = g^{\gamma\delta}$.  We also used
 \begin{align}
  G_{MN} J^M{}_\mu J^N{}_\nu = G_{\mu\nu} \,, \quad
  G_{MN} J^M{}_m J^N{}_n = G_{mn} \,, \quad
  G_{MN} J^M{}_i J^N{}_j = 0 \,,
 \end{align}
which is to say that $J^{MK} J^N{}_L$ acts as a projection in the tangent plane onto the $(X^\mu,Y^m)$ directions.  The last term in (\ref{ExpandOut}) is manifestly
positive, so ${\rm Area}(\Sigma) \ge \int_\Sigma J$, and $J$ is a
calibration.\footnote{See \cite{Joyce:2001xt} for an introduction
to calibrated surfaces.}  This inequality is saturated by minimal
surfaces calibrated with respect to $J$, which satisfy ${\cal
P}=0$ or $v_\alpha^M = 0$ (which includes $\partial_\alpha U^i
=0$). The vanishing of $v_\alpha^M$ defines a pseudo-holomorphic
curve with respect to the almost complex structure $J^M{}_N$.

The pseudo-holomorphicity equations simplify if one chooses the worldsheet coordinates $(\sigma^1,\sigma^2)$ so that $g_{\alpha\beta} = e^\phi \delta_{\alpha\beta}$ for some function $\phi(\sigma^1,\sigma^2)$.  Then, taking $\kappa=1$, the conditions $v^M_\alpha = 0$ boil down to
 \begin{align}
  &\partial_\alpha X^\mu - \epsilon_{\alpha\beta}{ R^2 \over Y^2 +
   U^2} \partial_\beta Y^m \delta^\mu_m =0 \label{xeqn}\\
   \label{pseudoholo2}
  &\partial_\alpha U^i = 0 \,.
 \end{align}
Since $dJ=0$ and $J$ obeys (\ref{jsq}), these equations are consistent, after using the
equations of motion for the surface.  Writing $Y^m = \frac{R^2}{Z}\hat\theta^m$ with $Z^2 = R^4/(Y^2 + U^2)$, \eqref{xeqn} becomes
 \begin{align}\label{xeqn2}
  \partial_\alpha X^\mu = \epsilon_{\alpha\beta}
    (Z\partial_\beta\hat\theta^m - \hat\theta^m \partial_\beta Z)
     \delta_m^\mu \,.
 \end{align}
For a curve which behaves smoothly at the
$Z=\epsilon\rightarrow 0$  boundary, equation \eqref{xeqn2}
implies the boundary condition $\dot x^\mu(s)/|\dot x| =
\delta^\mu_m \hat \theta^m(s)$, corresponding to a supersymmetric
Wilson loop.  The area of the curve is given by
 \begin{align}
  {\rm Area}(\Sigma) = \int_\Sigma J = \int_\Sigma \delta_{\mu m}
  d(Y^m dX^\mu) = \frac{1}{\epsilon}\int_{\partial
   \Sigma}\delta_{\mu m} \hat\theta^m dX^\mu= \frac{1}{\epsilon}
   \int ds |\dot x| \,.
 \end{align}
Thus, the finite part of the area vanishes.

Let's check that the minimal surface for the circular
supersymmetric Wilson loop found by Zarembo \cite{Zarembo:2002an},
is a pseudo-holomorphic curve.  This surface is a map of the disk of radius $\rho$ to $AdS_5 \times S^5$.  It is convenient to parametrize the disk with $(z,\varphi)$, where $z^2 = \rho^2-r^2$ and $r \in (0,\rho)$ is the usual radial variable for the disk.  Then the map to $AdS_5 \times S^5$ is of the following form:
 \begin{align}\seqalign{\span\TL & \span\TR \qquad & \span\TL & \span\TR \qquad & \span\TL & \span\TR}{
  X^1 &= \sqrt{\rho^2-z^2} \cos\varphi &
  X^2 &= \sqrt{\rho^2-z^2} \sin\varphi &
  X^3 &= X^4 = 0  \cr
  Y^1 &= -{R^2 \over \rho} {\sqrt{\rho^2-z^2} \over z} \sin\varphi &
  Y^2 &= -{R^2 \over \rho} {\sqrt{\rho^2-z^2} \over z} \cos\varphi &
  Y^3 &= Y^4 = 0  \cr
  U^1 &= {R^2 \over \rho} &
  U^2 &= 0 \,.
 } \label{circularsurface} \end{align}
To check that indeed $v^\alpha_M = 0$, it helps to make the following observations:
 \begin{align}\label{observations}\seqalign{\span\TL & \span\TR \qquad & \span\TL & \span\TR}{
  Z &\equiv R^4/(Y^2+U^2) = z &
  ds_\Sigma^2 &= R^2 \left( 1 + {\rho^2 \over z^2} \right)
   \left( {dz^2 \over \rho^2-z^2} + {\rho^2-z^2 \over \rho^2}
     d\varphi^2 \right)  \cr
  \jmath_{z\phi} &=
   {R^2 \over \rho} \left( 1 + {\rho^2 \over z^2} \right) &
  \jmath^z{}_\phi &= {1 \over \jmath_z{}^\phi} =
    {\rho \over \rho^2-z^2} \,.
 }\end{align}
It can also be checked that the area element on $\Sigma$ is precisely the pull-back of $J$:
 \begin{align}\label{Jagrees}
  dA|_\Sigma = J|_\Sigma =
    {R^2 \over \rho} \left( 1 + {\rho^2 \over z^2} \right)
      dz \wedge d\varphi \,,
 \end{align}
as expected since $J$ calibrates the surface.

\subsection{ More general cases}

If we consider a Dp-brane, we have a $p+1$-dimensional worldvolume
and we can consider a Wilson surface lying in an $n$-dimensional
subspace, $n \leq p+1$. We can then select an $n$-dimensional
subspace of the transverse space and form the almost complex
structure as above, from the exact two form $J = dX^m \wedge
dY^m$. We can then consider supersymmetric Wilson loops living in
the $n$-dimensional subspace and their corresponding
pseudo-holomorphic  surfaces in the gravity dual. In addition, we
can consider the Dp-brane theories in the Coulomb branch, whose
associated supergravity solutions are characterized by the
harmonic functions
\begin{equation}
f = \sum_i { N_i \over |\vec r - \vec r_i |^{ 7-p} } \,.
\end{equation}
Then the BPS  condition for the surface  becomes
 \begin{align}
  &\partial_\alpha X^\mu - \epsilon_{\alpha\beta} f^{1/2}
  \partial_\beta Y^m \delta^\mu_m =0 \label{xeqn3}\\
  \label{pseudoholo3}
  &\partial_\alpha U^i = 0 \,,
 \end{align}
where we have split the
transverse coordinates into $Y^m$ and $U^i$.

We can also consider these twisted theories  on a general four
manifolds. The Wilson loop operators are operators in these
theories. In principle we could find the gravity dual of these
field theories. Some simple example of topologically twisted
theories were considered in \cite{JMCN,gaunt}.  Although we have not made any explicit checks, we expect that
these geometries will also have an almost complex structure, so
that there are pseudo-holomorphic surfaces corresponding to
the BPS Wilson loops.

Another interesting case to consider is that of topological
strings, for a review see \cite{ancv}.  For example, in the
topological A model one specifies a symplectic form $J$ and, by
picking a metric, one can consider pseudo-holomorphic maps with
respect to $J_i^j$. These are the surfaces that contribute to the
topological A model. In this model one can consider the so called
A-branes
  which wrap Lagrangian submanifolds.
 In some cases\footnote{This happens when there are no worldsheet instantons.}
  the open string theory living on them
is a Chern-Simons theory. A particular case, studied in
\cite{rgcv} involves D3 branes wrapping the $S^3$ in the six-dimensional manifold $T^*(S^3)$, which  is the deformed conifold.
There it was conjectured that this theory is large $N$ dual to the
topological string theory on the deformed conifold, with
deformation parameter $t = 2 \pi i N g^2_{CS}$, where $g^2_{CS} =
1/(k + N)$ is the renormalized Chern-Simons coupling. Wilson loops
were considered in \cite{hocv} by introducing additional D-branes.
Here we simply point out that the natural Wilson loops to consider
are the supersymmetric Wilson loops discussed in this article. The
corresponding Wilson surfaces correspond to surfaces that lie
along the contour $ {\cal C}$ on $S^3$ and are extended along the
cotangent direction given by multiplying the tangent vector in
${\cal C}$ with $J_i^{\, j}$. According to the conjecture in
\cite{rgcv} the large $N$ results for these Wilson loops can be
obtained by considering topological strings in the resolved
conifold. The resolved conifold has a non-trivial $S^2$ and there
are  surfaces that wrap this $S^2$. In this case the $S^2$ can be
wrapped in two ways (see \cite{hocv}) and the genus zero answer
is\footnote{ The all genus answer can be found in formula (3.13)
of \cite{hocv}. See also \cite{wittenjones}.}
\begin{equation} \label{wtop}
\langle W \rangle_0 = { e^{ t / 2} - e^{- {t / 2 }} } \,,
\end{equation}
where $t$ is a parameter measuring the (complex) volume of $S^2$.
The two powers in (\ref{wtop}) arise from a surface that does not
wrap the $S^2$ and from a surface that wraps it once. In appendix B,
this is discussed in more detail. We see that in this case, where we
have a non-trivial 2 cycle in the geometry, we can have a more
complicated answer.  In contrast with $AdS^5 \times S^5$, $J$ is not
exact for the resolved conifold.

\section{Existence and counting of solutions}
\label{COUNTING}

Thus far we have shown that pseudo-holomorphic curves with smooth
behavior at the boundary of $AdS_5 \times S^5$ are minimal
surfaces corresponding to supersymmetric Wilson loops and have
vanishing regularized area.  We will now consider the converse
question of whether a pseudo-holomorphic curve exists for any
supersymmetric boundary condition, and whether there is a
non-trivial moduli space of solutions.

The pseudo-holomorphicity equation \eqref{xeqn} implies
\begin{align}\label{2ndord} \partial_\alpha
 \left( \frac{1}{Y^2 + U^2}\partial_\alpha Y^m \right) =0 \,.
\end{align}
It is again convenient to work with coordinates $Z$ and
$\hat\theta^A$, related to $Y^m$, $U^i$ by \\$Z^2 = 1/(Y^2 + U^2),$
$Y^m = \frac{1}{Z}\hat\theta^m$, $U^{1,2} = \frac{1}{Z} \hat
\theta^{5,6}$. In these coordinates, \eqref{2ndord} becomes
\begin{align}
Z\partial^2 \hat\theta^m - \hat\theta^m\partial^2 Z =0 ,\,\,\,
m=1\cdots 4 \,,
\end{align}
while \eqref{pseudoholo2} becomes
 \begin{align}\label{constant}
  \partial_\alpha \left( \frac{1}{Z}\hat\theta^{5,6} \right)
  =0\,.
 \end{align}
We may always do a $U(1)$ rotation in the $5,6$ planes such that
$\hat\theta^6 = 0$. Henceforward, we will take $\hat\theta^A$ to
indicate $\hat\theta^{1\cdots 5}$.  By defining $\lambda \equiv
\partial^2 Z / Z$ we may re-express \eno{2ndord} as
 \begin{align}\label{tp}
  &\partial^2 \hat\theta^m - \lambda\hat\theta^m = 0,\,\,\,
    m=1\cdots 4 \nonumber\\
  &\partial^2 Z - \lambda Z =0 \,,
 \end{align}
where we have used $\hat\theta^A\hat\theta^A=1$.  If $\hat\theta^5$ does
not vanish everywhere, \eqref{constant} implies $Z= c\theta^5$ for
constant $c$, so the last equation of \eqref{tp} gives $\partial^2
\theta^5 - \lambda \hat\theta^5 =0$.   Thus
\begin{align}\partial^2 \hat\theta^A - \lambda\hat\theta^A = 0,
\,\,\, A =1\cdots 5 \,. \label{geneq} \end{align} These are the
equations of an $S^4$ sigma model,  with $\lambda$ playing the
role of a Lagrange multiplier enforcing the condition
$\hat\theta^A\hat\theta^A =1$.  These equations must be solved with
boundary values $\hat\theta^{1\cdots4}$ given by the Wilson loop
parameters $\hat\theta^m(s)$,  and $\hat\theta^5 = Z =0$.  A simple example of a solution is provided by the minimal surface found by Zarembo for the circular Wilson loop \eno{circularsurface}:
\begin{align}\label{Zarsoln} &\hat\theta^1 = - \sqrt{1-Z^2}\sin\varphi,\,\, \hat\theta^2 =
\sqrt{1-Z^2}\cos\varphi,\nonumber
\\ &\hat\theta^3=\hat\theta^4=\hat\theta^6 =0,\\ &\hat\theta^5 =
Z \,.\nonumber
\end{align}

We would like to show that given any contour, $X^\mu(\tau)$, there exists a solution to \eno{tp} which matches onto it asymptotically.  We have not been able to do this, but let us give a plausibility argument for the existence of such a solution.  Given the contour, we can certainly find $\hat \theta_B^m = \delta^m_\mu d_s X^\mu(\tau)$,
where $ds = |d_\tau X^\mu|$ is the proper length element
 along
the contour.  Now we imagine an ansatz where the worldsheet, in
conformal gauge,  has the topology of the disk, parametrized in
terms of $r$ and $\varphi$, where $\varphi$ is the angular
variable.  We suppose that $\varphi = \varphi(\tau)$.
In terms of the unknown function $\varphi(\tau)$ we can write
the boundary conditions for $\hat \theta^A$ on the boundary of the
disk.  It is clear that, given the sigma model equations, we will
find a solution to these equations with these boundary values and
$\hat \theta^5 \geq 0$.  We then we set $Z = c \hat \theta^5$.  At
this point we can write down the equation for $\varphi(\tau)$:
 \be
 c \partial_r \hat \theta^5
 |_{r=1} = | \partial_\varphi X^i|_{r=1} =  |\partial_\tau X^\mu|
 { d \tau \over d \varphi} \,.
 \ee
This is appears to be a very complicated equation for $\varphi(\tau)$.  But it is one functional equation for one function, so we expect that it should have a solution.  We have explicitly checked that for arbitrary small deformations of the circular contour in ${\bf R}^2$, the solution exists to first order.

In the case that the contour is in ${\bf R}^2$, and the boundary
conditions for $\hat \theta^{1,2}$ wind once around a great circle on $S^5$, then we can compute a formula for the value of the constant $c$ that relates $\hat \theta^5$ and $Z$.  In this case the worldsheet will cover the upper hemisphere of $S^2$. We can consider the $SO(3)$ currents $J^{ij} = \hat\theta^i d\hat\theta^j - \hat\theta^j d\hat\theta^i$, where $i$ and $j$ run over $1,2,5$.  Using $Z = c
\hat\theta^5$, we rewrite (\ref{xeqn2}) as
\begin{equation}
d X^i = * ( Z d \hat\theta^i - d Z \hat\theta^i) = c * J^{5 i}
~,~~~~~~~~i=1,2 \,.
\end{equation}
Let us now consider the expression for the area\footnote{ Do not confuse
this area, which is computed using the four-dimensional flat
metric of the boundary theory, with the area of the surface in ten
dimensions.} enclosed by the contour:
 \begin{equation}
 A_{\rm FT} = \int dX^1 \wedge dX^2 = - c^2 \int J^{5 1} \wedge J^{5 2}  =
  { 1 \over 2 i }   \int J^{5 +} \wedge J^{5 -}
 \end{equation}
where  we defined $J^{5 \pm } = J^{51} \pm i J^{5 2} $.  With the standard parametrization of $S^2$ as $\hat\theta^5 = \cos\beta$ and $\hat\theta^1 + i \hat\theta^2 = e^{i \alpha} \sin \beta$, we find that
  \begin{equation}
  J^{5 \pm } = e^{ \pm i \alpha} \left[ d \beta \pm i  \cos \beta \sin
  \beta \, d \alpha  \right] ~,~~~~~~~ { 1\over 2 i } \int J^{5 +}
  J^{5-} = \pi \,,
  \end{equation}
where we used that the worldsheet wraps half of the sphere.
  This implies that
  \begin{equation} \label{arearel}
   { A_{\rm FT} \over \pi } = c^2 \,.
  \end{equation}
Note that the functional dependence in this relation is determined
by conformal symmetry. On the other hand, it is interesting that
the numerical coefficient  is independent of the shape of the
contour.

The solutions are not unique in general.  In fact the circular
solution (\ref{circularsurface}) has a moduli space of solutions
since we can rotate it in the directions $(Y^3,Y^4,U^1,U^2)$, so
the moduli space of solutions is an $S^3= SO(4)/SO(3)$. Notice
that this solution has the feature that it breaks spontaneously
the $SO(4)$ symmetry that the Wilson loop operator preserves. Of
course, this $SO(4)$ symmetry is restored after we integrate over
these moduli.  The same symmetry breaking pattern occurs for other
planar Wilson loops, and we expect a similar phenomenon to arise for
 generic Wilson loops. Namely,  we expect that for $1/8$
supersymmetric Wilson loops, with boundary conditions
$\theta^{4,5,6}=0$, there is an SO(3) symmetry which preserves the
boundary conditions.  Acting with SO(3) on a minimal surface with
$Z=c \theta^5 $ shows that there is at least an $SO(3)/SO(2) = S^2$
moduli space of solutions. Finally, for a generic $1/16$ BPS surface
we expect an $SO(2)$ that acts on the solution and the moduli space
would be at least an $S^1$.  We do not know if there are other
solutions.  In light of \eqref{geneq}, the counting of solutions is
related to the counting problem for harmonic maps of a disc to $S^n$
with Dirichlet boundary conditions.  This is a difficult problem,
for which only partial results are known
\cite{BrezisCoron,BenciCoron,Brezis}.

Note that the non-renormalization theorem in \cite{GK} implies that
the expectation value of the Wilson loop is exactly one, so all
$\alpha'$ (or $1/\sqrt{\lambda}$) corrections should vanish.
Moreover, the leading term in the loop expansion \eqref{semicl}
should not give any dependence on $\sqrt{\lambda}$,  suggesting
\cite{Zarembo:2002an} that the number of zero modes should be three.
We have not understood the resolution of this apparent discrepancy
between field theory and string theory calculations.  It is possible
that we have under-counted the number of collective coordinates in
the cases with less than $1/4$ supersymmetry.

\section{Supersymmetry preserved by the surface}
\label{SUSY}

In this section we show that a surface obeying (\ref{pseudoholo2}) preserves some supersymmetry.

Let us first consider a string worldsheet in flat space extended along directions $\hat 1$ and
$\hat 2$. This worlsheet will preserve supersymmetries that obey
$\Gamma_{12} \epsilon_L = i \epsilon_L $, $\Gamma_{12}\epsilon_R = - i \epsilon_R$,
where $\epsilon_{L,R}$ are the two spinors of type
II string theory. Now consider a worldsheet embedded in a more general spacetime.
Then, at each point on the surface, we have a condition which is the same as the one we
had in flat space. We can write this as
\begin{equation} \label{condgen}
\epsilon^{\alpha \beta} \partial_\alpha X^M \partial_\beta X^N
\Gamma_{MN} \epsilon_{L,R} = \pm i G_{MN} \partial_\alpha X^M
\partial_\alpha X^N \epsilon_{L,R} \,,
\end{equation}
where the $\pm$ is correlated with $L, R$.

The AdS Poincar\'e supersymmetries are generated by spinors of the
form
\begin{eqnarray}
\epsilon_{L,R} &=& (Y^2 + U^2)^{-{1 \over 4} } \epsilon_{L,R \, 0}
\label{condspinone} \\
i \hat \Gamma_{1234} \epsilon_{L\, 0 } &=& \epsilon_{R \,  0} \,,
\label{condspintwo}
\end{eqnarray}
where $\epsilon_{l,R \, 0}$ are flat space spinors and $\hat
\Gamma $ are flat space gamma matrices. We have written the
equation in Euclidean space, a fact which is responsible for the
extra $i$ in (\ref{condspintwo}). We can view equation
(\ref{condspintwo})  as giving $\epsilon_R$ once we have
$\epsilon_L$.

Suppose now that we have a general pseudo-holomorphic Wilson
surface. We then choose a spinor $\epsilon_{L \, 0}$ which obeys
\begin{equation}
\Gamma_{MN} \epsilon_{L } = i J_{M N} \epsilon_{L }
 ~,~~~~{\rm or} ~~ \hat \Gamma_{X^j Y^j} \, \epsilon_{L \, 0} = i \, \epsilon_{L \, 0} ~,~~
 ({\rm no~sum~over~}j) ~,~~~~~~
\label{condwant}
\end{equation}
  These conditions imply that
 \begin{equation} \label{twistcond}
 (\Gamma_{X^i X^j} + \Gamma_{Y^i Y^j} )\epsilon_{L \, 0 } =0 \,,
 \end{equation}
 which is the condition that the spinor is annihilated by the twisted
spin connection. So the spinor we are considering is the one
related to the BRST operator of the twisted theory.\footnote{ There
are two spinors that obey (\ref{twistcond}), which obey $\hat
\Gamma_{MN} \epsilon_{L \, 0} = \pm i J_{M N} \epsilon_{L \, 0}$.
We are interested in only one of them.}
    We can now show that
\begin{eqnarray}
&& \left( \epsilon^{\alpha \beta} \partial_\alpha X^M
\partial_\beta X^M \Gamma_{MN}  - i G_{MN} \partial_\alpha X^M
\partial_\alpha X^M \right) \epsilon_{L\, 0}
 =
\\ &&  i \left(\epsilon^{\alpha \beta} \partial_\alpha X^M \partial_\beta
X^M J_{MN}  - G_{MN} \partial_\alpha X^M \partial_\alpha X^M
\right) \epsilon_{L \, 0}  = \\
&& - { i \over 2} G_{MN} v_{\alpha}^M v_\alpha^N \epsilon_{L \, 0}
=0 \,, \label{susyspv}
\end{eqnarray}
where we used that pseudo-holomorphic surfaces are defined by the
condition that $v_\alpha^M =0$, see (\ref{ThisPos}).

Notice that if we define $\epsilon_{R}$ through
(\ref{condspintwo}) we then get that $\epsilon_{R \, 0}$ obeys
$\Gamma_{MN} \epsilon_{R \, 0 } = - i J_{MN} \epsilon_{R \, 0}$.
This extra minus sign cancels the extra sign in (\ref{condgen}),
so that we end up with the same combination $v_\alpha^M$ which
appeared in (\ref{susyspv}), which vanishes for a
pseudo-holomorphic surface.

\section{Pseudo-holomorphic surfaces with a $U(1)$ isometry}
\label{ANNULUS}

As we have discussed in section~\ref{COUNTING}, a
pseudo-holomorphic surface in $AdS_5 \times S^5$ should exist with
any given supersymmetric Wilson loop specifying its boundary data.
However, to find explicit examples, it is easier to start
with a surface and extract the Wilson loop.  More precisely, one
may start with a solution to the equations (see \eno{geneq})
 \eqn{FiveEQN}{
  \partial^2 \hat\theta^A - \lambda\hat\theta^A = 0 \,, \quad
   \sum_A \hat\theta^A \hat\theta^A = 1 \,, \qquad
   A = 1\cdots5 \,,
 }
with $\hat\theta^5 > 0$ somewhere, and work backwards to the surface and the Wilson loop.  We will employ coordinates $(t,\varphi)$ which are assumed to lead to a metric in conformal gauge, so that \eno{FiveEQN} applies as written.  The procedure we will follow in order to obtain a Wilson loop is:
 \begin{enumerate}
  \item Choose a solution to \eno{FiveEQN}.
  \item Choose a maximal connected region $R$ in the space of coordinates $(t,\varphi)$ on which $\hat\theta^5 > 0$.
  \item Extract $Y^m$ and $U^i$ by setting $Z = c \hat\theta^5$, $Y^m = \hat\theta^m/Z$, and $U^i = (1/c,0)$, where $c$ is an arbitrary constant that sets the overall scale of the Wilson loop.
  \item Use the pseudo-holomorphicity equations to determine $X^\mu$.
 \end{enumerate}

As remarked previously, \eno{FiveEQN} are the equations of motion of an $S^4$ non-linear sigma model with action
 \eqn{SigmaAction}{
  S = \int dt d\varphi \,
    {1 \over 2} \left[ (\partial\hat\theta^A)^2 +
    \lambda (\hat\theta^A \hat\theta^A - 1) \right] \,.
 }
It is clearly difficult to characterize all the classical solutions in this theory.
 In the following sections, we will address two special cases which include the circular
  Wilson loop of \cite{Zarembo:2002an}.  The simplifying feature in both cases is an
  abelian isometry which
allows us to reduce the problem to classical dynamics of one or two degrees of freedom.
 The Wilson loops in question are special cases of a general class studied in
 \cite{DrukkerFiol}, and we will make use of some of the analytical methods
 developed there.  We also explain in section~\ref{SECONDANSATZ} a connection between
 the Wilson loop problem and Poincar\'e's stroboscopic map.

\subsection{The first ansatz}
\label{FIRSTANSATZ}

The first approach we will explore is to restrict to
 \eqn{FirstAnsatz}{
  \hat\theta^1 + i \hat\theta^2 = e^{i\varphi} \cos\eta(t) \,, \qquad
   \vec\theta
    = \begin{pmatrix}
       \hat\theta^3 \cr \hat\theta^4 \cr \hat\theta^5
      \end{pmatrix}
    = \vec\theta(t)
    = \sin \eta(t) \begin{pmatrix}
       \sin\alpha(t) \cos\beta(t) \cr \sin\alpha(t) \sin\beta(t) \cr
        \cos\alpha(t)
      \end{pmatrix} \,,
 }
which leads to
 \eqn{SigmaReduced}{\eqalign{
  {S \over 2\pi} &= \int dt \, L \,, \qquad L = {1 \over 2} \left[
    \dot\eta^2 + \sin^2 \eta \, (\dot\alpha^2 + \sin^2 \alpha \,
      \dot\beta^2) + \cos^2 \eta \right]  \cr
  H &= {p_\eta^2 \over 2} + {p_\alpha^2 \over 2 \sin^2 \eta} +
   {p_\beta^2 \over 2\sin^2 \eta \sin^2 \alpha} -
   {1 \over 2} \cos^2 \eta = E  \cr
  p_\eta &= \dot\eta \,,\quad
    p_\alpha = \sin^2 \eta \, \dot\alpha \,,\quad
    p_\beta = \sin^2 \eta \sin^2 \alpha \, \dot\beta \,.
}} The simplest case to consider is where there is no ``angular
momentum'' in the $\alpha$, $\beta$ directions.  This means that
$\vec\theta(t) = \sin\eta(t) \, \hat{n}$ for a fixed vector unit
vector $\hat{n}$.  Without loss of generality we set  $\hat{n} =
(0,0,1)$.
 In terms of the variable $ 2 \eta$ is is clear that we have the
 equation of motion for a simple pendulum.
Once we have a general solution we can straightforwardly obtain
 \eqn{GotYs}{\eqalign{
  Z &= \sin \eta  \cr
  Y^m &= {1 \over Z} \left( \cos \eta
    \cos\varphi, \cos \eta \sin \varphi, 0, 0 \right)  \cr
  X^\mu &= \left( {\dot\eta } \sin\varphi,
   -{\dot\eta } \cos\varphi, 0, 0 \right) \,,
 }}
where we have set $c=1$.  Now, $\dot\eta$ may be expressed in
terms of the total energy:\footnote{ $\epsilon^2 = 2 E +1$, where
$E$ is the energy in (\ref{SigmaReduced}).}
 \eqn{DotPsi}{
  {\dot \eta }  = \sqrt{\epsilon^2 -\sin^2 \eta} \,.
 }
 Putting \eno{GotYs} and \eno{DotPsi} together,
 we have a fully explicit parametrization of a surface.
 $X^\mu(s)$ for the associated Wilson loop can be determined by
 setting $Z=0$: it is in all cases a circle with radius $\epsilon$.
 Note that in this case the relation (\ref{arearel}) does not hold
 because the worldsheet does not cover the hemisphere once.

We should distinguish three cases depending on whether the energy
of the simple pendulum is equal, greater than, or less than than the maximum potential energy.
 \begin{enumerate}
  \item $\epsilon=1$. This describes the
   circular Wilson loop of \cite{Zarembo:2002an}.
  \item $\epsilon > 1$.
  The topology of the surface in $AdS_5 \times S^5$ is
  best described as an annulus with both boundaries lying on the boundary
   circle.
  To see this, consider the curve with $\varphi=\pi/2$.
  It starts at $\eta=0$ at the point $X^\mu = \left( \epsilon,0,0,0 \right)$
  of the
  boundary circle, then goes inward until $\eta={\pi \over
2}$, where $X^\mu =
  \left( \sqrt{\epsilon^2-1},0,0,0 \right)$.  From this central point, the
  curve proceeds back outwards until $\eta= \pi$, where it returns to the point
  $X^\mu = \left( \epsilon,0,0,0 \right)$.  Meanwhile, $Y^m$ goes from $(0,+\infty,0,0)$ to $(0,0,0,0)$ to $(0,-\infty,0,0)$.

Evidently, this annulus describes the connected correlator of two
Wilson loops whose  real space parts are identical but whose
scalars are equal and opposite.  Section~\ref{COINCIDENT} includes further discussion of these correlators.
  \item $\epsilon < 1$.  Again restricting to $\eta=0$, the curve \eno{GotYs}
  is the real space description of the boundary.
   But as before, the topology of the surface in
   $AdS_5 \times S^5$ is an annulus.
   A curve with $\varphi=\pi/2$ starts at $\eta=0$, goes
   to $\eta=\eta_{\rm max} < {\pi \over 2}  $, and returns to $\eta=0$.
    Meanwhile, $X^\mu$ goes from
    $\left( \epsilon,0,0,0 \right)$
    to $(0,0,0,0)$ to $\left( -\epsilon,0,0,0 \right)$,
     and $Y^m$ goes from $(0,+\infty,0,0)$ to $(0,\cot \eta_{\rm max} ,0,0)$
     to $(0,+\infty,0,0)$.

Again we are describing the connected correlator of two Wilson loops whose real space parts are identical but whose scalars are equal and opposite, but the connectivity of points on one loop to points on the other is different from the previous case.
 \end{enumerate}
If $\epsilon$ is adjusted continuously to $1$, the result is two
coincident but unconnected circular Wilson loops.  In the limit of
small energy, the projection of the shape onto $AdS_5$ is similar
to two copies of the shape for a non-BPS circular Wilson loop
\cite{Drukker:1999zq}.

A qualitatively different situation arises when there is angular momentum in the directions parameterized by $\alpha$ and $\beta$ in \eno{SigmaReduced}.  By performing a rigid $SO(3)$ rotation on $\vec\theta$ if necessary, it is possible to choose $\alpha(t) \equiv \pi/2$: this is essentially the statement that central force motion occurs in a plane that includes the origin.  Conservation of the angular momentum
 \eqn{pBetaDef}{
  p_\beta = {\partial L \over \partial \dot\beta}
    = \sin^2 \eta \, \dot\beta
 }
and the total energy $E$ appearing in \eno{SigmaReduced} makes it possible to extract the solution in integral form: assuming $\dot\eta \neq 0$,
 \eqn{tSolve}{
  t = \int {d\eta \over \sqrt{2E - p_\beta^2/\sin^2 \eta +
    \cos^2 \eta}} \,.
 }
The integral can be performed in terms of elliptic functions.

From a plot of the effective potential, one can see that there are several qualitatively different possible behaviors.
 \begin{figure}
 \begin{center}
  \includegraphics{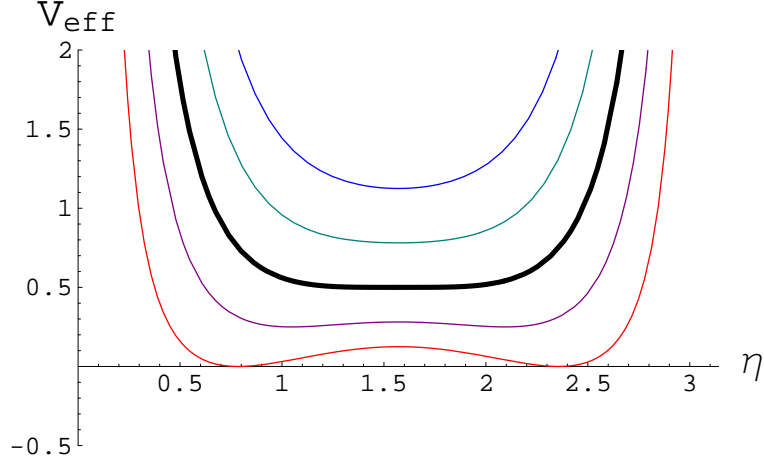}
  \caption{$V_{\rm eff}(\eta)$ versus $\eta$ for different values of $p_\beta$.  The dark curve has $p_\beta = 1$.  The ones above it have $p_\beta > 1$, and the ones below it have $0 < p_\beta < 1$.}\label{Veff}
 \end{center}
 \end{figure}
A common feature, however, is that $\eta$ can never equal $0$ or $\pi$, so $\hat\theta^5 = \sin\eta \neq 0$.  This must be remedied by sending $\hat\theta^A \to O^A{}_B \hat\theta^B$ for some $SO(5)$ rotation, which we take to be of the form
 \eqn{InterestingO}{
  O = \begin{pmatrix} \cos\gamma & 0 & 0 & 0 & -\sin\gamma \cr
    0 & 1 & 0 & 0 & 0 \cr
    0 & 0 & 1 & 0 & 0 \cr
    0 & 0 & 0 & 1 & 0 \cr
    \sin\gamma & 0 & 0 & 0 & \cos\gamma
   \end{pmatrix} \,.
 }
To focus the discussion, we restrict ourselves further to the simplest case where
$\eta(t)=\eta_0$, a constant.
 In order to have $V_{\rm eff}'(\eta) = 0$, we must either set
 $\eta_0 = \pi/2$ or, for $0 < p_\beta < 1$, choose $\sin^2\eta_0
 =p_\beta
   $.  The latter choice is the one that we will treat here.
  Then one straightforwardly obtains
 \eqn{FancyCircle}{\eqalign{
  X^\mu &= \sin\gamma \, \left( 0, (1-p_\beta) t,
    \sqrt{p_\beta(1-p_\beta)} \sin\varphi \sin t,
    -\sqrt{p_\beta(1-p_\beta)} \sin\varphi \cos t \right)  \cr
  Y^m &= \csc\gamma \, \left( \cos\gamma, \tan\varphi,
    \sqrt{p_\beta \over 1-p_\beta} {\cos t \over \cos\varphi},
    \sqrt{p_\beta \over 1-p_\beta} {\sin t \over \cos\varphi } \right)  \cr
  Z &= \sqrt{1-p_\beta} \cos\varphi \sin\gamma \,.
 }}
The topology is a strip with boundaries at $\varphi = \pm\pi/2$ ending on the helices
 \eqn{SimpleCircle}{
  X^\mu = \sin\gamma \,
   \left( 0,(1-p_\beta) t,\pm\sqrt{p_\beta(1-p_\beta)} \sin t,
    \mp\sqrt{p_\beta(1-p_\beta)} \cos t \right) \,.
 }
A double-helix Wilson loop also appears in \cite{DrukkerFiol}.

\subsection{The second ansatz}
\label{SECONDANSATZ}

Now let us consider another obvious ansatz involving an abelian isometry:
 \eqn{OtherAnsatz}{\seqalign{\span\TL & \span\TR &\qquad \span\TL & \span\TR &\qquad \span\TL & \span\TR}{
  \hat\theta^1 + i \hat\theta^2 &= \rho_1(t) e^{i\varphi} &
  \hat\theta^3 + i \hat\theta^4 &= \rho_2(t) e^{ib\varphi} &
  \hat\theta^5 &= \rho_3(t)  \cr
  \rho_1(t) &= \cos\eta(t) \cos\xi(t) &
  \rho_2(t) &= \cos\eta(t) \sin\xi(t) &
  \rho_3(t) &= \sin\eta(t) \,,
 }}
where $b$ is an arbitrary real number greater than $1$.  Unless $b$ is rational, $\varphi$ is
not a periodic coordinate.  Nevertheless, correct equations of motion
can be extracted from a reduced action,
 \eqn{Sreduced}{
  S_{\rm reduced} = \int dt \, L \,, \qquad
   L = {1 \over 2} \left[ \dot\eta^2 + \cos^2 \eta \, \dot\xi^2 +
     \cos^2 \eta (\cos^2 \xi + b^2 \sin^2 \xi) \right] \,.
 }
This system is integrable \cite{DrukkerFiol,russo}, with conserved
quantities $I_1$, $I_2$, and $I_3$ defined by
 \eqn{IDefs}{
  I_i = \rho_i^2 - \sum_{j \neq i}
    {(\rho_i \dot\rho_j - \rho_j \dot\rho_i)^2 \over m_i^2-m_j^2} \qquad
  m_1 = 1 \,,\quad m_2 = b \,,\quad m_3 = 0 \,,
 }
and subject to $I_1+I_2+I_3=1$.  The total energy may be expressed as $H = -(I_1+b^2 I_2)/2$.  Because of these two linear relations, the two independent conserved quantities can be chosen as $I_1$, $I_2$, or as $H$, $I_3$.

To integrate the system explicitly, it is helpful to use elliptical coordinates $\zeta_1$, $\zeta_2$, defined equivalently as the solutions of
 \eqn{ZetaDefOne}{
  {\rho_1^2 \over \zeta - 1} + {\rho_2^2 \over \zeta - b^2} +
    {\rho_3^2 \over \zeta} = 0 \,,
 }
or through
 \eqn[c]{ZetaDefs}{
  \sin\eta = {\sqrt{\zeta_1 \zeta_2} \over b} \qquad
  \tan\xi = {1 \over b} \sqrt{-{(b^2-\zeta_1)(b^2-\zeta_2) \over
   (1-\zeta_1)(1-\zeta_2)}} \,,
 }
and subject to the inequalities
 \eqn{ZetaInequalities}{
  0 \leq \zeta_1 \leq 1 \leq \zeta_2 \leq b^2 \,.
 }
It is straightforward to show that
 \eqn{ZetaSolve}{\eqalign{
  \dot\zeta_1^2 &= {f(\zeta_1) \over (\zeta_1-\zeta_2)^2} \qquad
   \dot\zeta_2^2 = {f(\zeta_2) \over (\zeta_1-\zeta_2)^2}  \cr
  f(\zeta) &\equiv 4 \zeta (b^2-\zeta)(1-\zeta)
   \left[ b^2 I_3 - (2H +1+b^2) \zeta +
    \zeta^2 \right]  \cr
    &\equiv 4 \zeta (b^2-\zeta)(1-\zeta)(\zeta_a-\zeta)(\zeta_b-\zeta)
      \,.
 }}
Note that $f(\zeta_1)$ and $f(\zeta_2)$ must be nonnegative on the range of values of $\zeta_1$, $\zeta_2$ that occur in a given solution.  This puts constraints on the values of $\zeta_a$ and $\zeta_b$.  Assuming that no $SO(5)$ rotation is applied, the boundary of the Wilson loop is at $\zeta_1 = 0$.  Therefore $f(\zeta_1) \geq 0$ for some closed interval whose left endpoint is $\zeta_1=0$.  The right endpoint must be where $f(\zeta_1)$ has a zero, namely at $1$ or $\zeta_a$, whichever is smallest.
Similar considerations of the possible range of $\zeta_2$ leads to the following possibilities for solutions:
 \def\min{\mop{min}}
 \def\max{\mop{max}}
 \eqn{SolutionIneqs}{\eqalign{
  0 &\leq \zeta_1 \leq \min\{1,\zeta_a\}  \cr
  \max\{1,\zeta_a\} &\leq \zeta_2 \leq \min\{b^2,\zeta_b\} \,.
 }}
The orbits of $\zeta_1$, $\zeta_2$ satisfy the integral equations
 \eqn{ZetaOrbits}{
  \int {d\zeta_1 \over \sqrt{f(\zeta_1)}} =
   \int {d\zeta_2 \over \sqrt{f(\zeta_2)}} \,.
 }

The explicit results \eno{ZetaSolve} and~\eno{ZetaOrbits} are not necessary to understand the qualitative features of supersymmetric Wilson loops associated with the ansatz \eno{OtherAnsatz}.  Consider a solution $\eta(t)$, $\xi(t)$ of the equations of motion following from \eno{Sreduced}, with $\eta(t) > 0$ on some finite or semi-finite interval.
Following the procedure outlined at the beginning of section~\ref{ANNULUS}, one obtains the following surface in $AdS_5 \times S^5$:
 \eqn{XYthetaSep}{\eqalign{
  X^\mu &= \dot\eta \, \left( \sin\varphi \cos\xi,
    -\cos\varphi \cos\xi, {1 \over b} \sin b\varphi \sin\xi,
    -{1 \over b} \cos b\varphi \sin\xi \right)  \cr
   &\qquad{}+ \dot\xi \cos\eta \sin\eta \left(
    \sin\varphi \sin\xi, -\cos\varphi \sin\xi,
    -{1 \over b} \sin b\varphi \cos\xi,
    {1 \over b} \cos b\varphi \cos\xi \right)  \cr
  Y^m &= \cot\eta \,
   (\cos\varphi \cos\xi, \sin\varphi \cos\xi,
    \cos b\varphi \sin\xi, \sin b\varphi \sin\xi)  \cr
  Z &= \sin\eta \,.
 }}
The real space part of the Wilson loop is
 \eqn{RealSpaceSep}{
  X^\mu = \dot\eta \, \left( \sin\varphi \cos\xi,
    -\cos\varphi \cos\xi, {1 \over b} \sin b\varphi \sin\xi,
    -{1 \over b} \cos b\varphi \sin\xi \right) \,,
 }
where now all quantities are evaluated at the time(s) when $\eta=0$.

It is worth noting an analogy with Poincar\'e's stroboscopic map.  In Poincar\'e's setup, one starts with a system with two degrees of freedom (such as $\eta$ and $\xi$), selects one as the ``timing'' variable ($\eta$, let's say), and then considers a map from the surface in phase space (the stroboscopic plane) defined by $\eta=0$, $\dot\eta > 0$, and $H=E$, where $E$ is constant.  The stroboscopic map is a volume-preserving bijection of the stroboscopic plane to itself defined by starting the system at a point on the stroboscopic plane, then evolving the system forward in time until it again meets the stroboscopic plane.

In the Wilson loop setup, $\eta=0$ is privileged once we choose $Z = c \hat\theta^5$.  (But, as we remarked above, other choices can be made through $SO(5)$ rotations.)  Energy conservation is also part of the Wilson loop setup.  But instead of a single stroboscopic plane, we should now consider two disjoint planes: $P_+$ defined by $\dot\eta > 0$ and $P_-$ defined by $\dot\eta < 0$ (with $\eta=0$ and $H=E$ in both cases).  There is then a natural map from $P_+$ to $P_-$ defined by taking a point on $P_+$ and evolving the system forward until it meets $P_-$.  This is ``half'' of the stroboscopic map: to complete it one would evolve forward from $P_-$ back to $P_+$.  A point on $P_+$ corresponds to one boundary of a Wilson loop correlator: \eno{RealSpaceSep} with $\dot\eta > 0$.  Such a curve can be fully specified by the data that selects a point on $P_+$: one can for example choose $\xi$, $\dot\xi$, and the total energy $E$ and solve for $\dot\eta > 0$.  Likewise a point on $P_-$ corresponds to a Wilson loop of the form \eno{RealSpaceSep} with $\dot\eta < 0$.

The evolution of the dynamical system from $P_+$ to $P_-$ traces out the string worldsheet in $AdS_5 \times S^5$.  Because the boundary has two disjoint parts, the configuration represents the correlator of two Wilson loops.  As in the example of the simple pendulum in section~\ref{FIRSTANSATZ}, it is also possible to have separatrix behavior, where for a particular choice of energy the system starts on $P_+$ and then evolves to infinite time without ever intersecting $P_-$.  This corresponds to a single Wilson loop rather than a correlator of two.

A more explicit description of single Wilson loop cases can be
obtained using the integrable structure and the elliptic
coordinates $\zeta_1$ and $\zeta_2$. Recall that the boundary of
the Wilson loop is at $\zeta_1=0$.  For the system to evolve to
infinite time without coming back to $\zeta_1=0$, it must
asymptote to some other value, call it $\zeta_*$. There must be a
double zero of $f(\zeta)$ at $\zeta=\zeta_*$, because otherwise
the system reaches $\zeta_1=\zeta_*$ in finite time. The only
possibility is $\zeta_*=\zeta_a=1$.  There must also be an
asymptotic value for $\zeta_2$, and there are two ways to arrange
this. One way is to choose $\zeta_b = b^2$, so that there is a
double root at $\zeta=b^2$ as well as at $\zeta=1$. Then,
arranging signs so that $\dot\zeta_2 > 0$, one obtains a
one-parameter family of smooth single-boundary surfaces by
choosing an arbitrary boundary value of $\zeta_2 \in (1,b^2)$. The
second way to get an asymptotic value for $\zeta_2$ at late time
is to make $\zeta_2$ constant for all time: choose
$\zeta_2=\zeta_b \in (1,b^2)$.

A Wilson loop with the coincident boundaries similar to what was
found in section~\ref{FIRSTANSATZ} may be constructed by setting a
double zero $\zeta_a=\zeta_b$ with $\zeta_2=\zeta_a$ for all
time.  Then $\zeta_1$ varies from $0$ to $1$ and the location of
the double zero $\zeta_a=\zeta_b$ is a parameter of the solution.
Alternatively, if $\zeta_2=\zeta_b=1$, then the location of $\zeta_a \in (0,1)$ is arbitrary
and $\zeta_1$ varies from $0$ to $\zeta_a$. This latter solution
corresponds to the physical pendulum $\eta(t)$ with $\xi=\pi/2$
and energy specified by $\zeta_a$.

\subsection{Coincident boundaries}
\label{COINCIDENT}

\begin{figure}[tb]
\begin{center}
\epsfxsize=4in\leavevmode\epsfbox{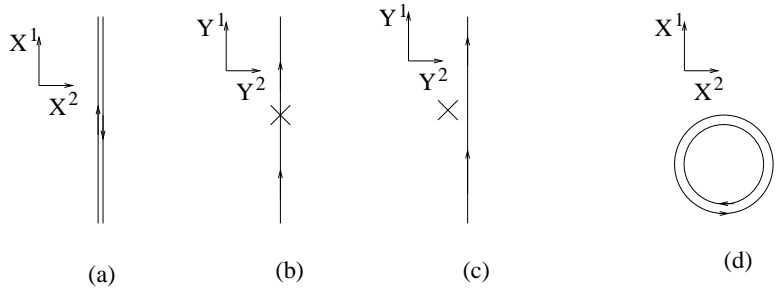}
\end{center}
\caption{  In (a) we see two coincident straight Wilson lines. We
have separated them to ease visualization, but they are on top of
each other. In (b) we plot two of the directions transverse to the
D-brane and we see a string ending on the D-brane (represented by
a cross) and a string leaving the D-brane. In (c) the two strings
from (b) combine and move in a direction transverse to the
D-brane. In (c) we see two coincident and oppositely oriented
circular Wilson loops. Again we have separated them just to ease
visualization.
 } \label{figone}
\end{figure}

A new feature of the annulus solutions discussed in
section~\ref{FIRSTANSATZ} and in the last paragraph of
section~\ref{SECONDANSATZ} is that two boundaries are coincident
but with opposite orientations. To simplify the discussion of this
situation, let us first suppose that we have a straight Wilson
loop and we now place an oppositely oriented Wilson line on top of
the first, see figure \ref{figone}(a). (This situation is in fact
the $E \to \infty$ limit of the annular solution discussed in
section~\ref{FIRSTANSATZ}).  The second Wilson line has the
opposite orientation, but it also has the opposite scalar charge,
so it preserves the same supersymmetry. In this case, the
corresponding surface in AdS has some zero modes. In other words,
there
 is a family
of surfaces that obey the boundary conditions. Let us  assume that
the Wilson lines are along $X^1$ on the D3-brane worldvolume. So
we can consider any surface that is extended along $X^1$ and $Y^1$
and sits at any point in $Y^2, \cdots , Y^6$. Thus the space of
zero modes is non-compact. We see that such surfaces obey the
equations (\ref{pseudoholo2}), for a suitable choice of worldsheet
coordinates. One way to understand the presence of these zero
modes is to think of the strings ending on a D3 brane. We have a
configuration with a string ending on the brane and one moving out
of the D3-brane, figure \ref{figone}(b). These two strings can be
joined and moved out of the D3 brane, see figure \ref{figone}(c).

Connected correlation functions of chiral Wilson loops vanish, via
the following standard argument. Supersymmetry implies that the
correlation function is independent of the separation.  Together
with clustering, this property gives a vanishing connected
correlator. From the AdS point of view,  the vanishing of the
connected correlation function suggests that the
pseudo-holomorphic annulus should have a fermionic zero-mode.
As a
point of comparison, recall that the pseudo-holomorphic disk is
not generically invariant under R-symmetries respected by the
boundary; for chiral Wilson loops in ${\bf R}^2$, the broken
R-symmetries give rise to bosonic zero modes.
All supersymmetries of the chiral Wilson loop should be respected
by the disk in order to get expectation value $1$ rather than $0$.
A counting of zero modes for the pseudo-holomorphic disk and
annulus, lacking at present, is required to give a complete check
of the non-renormalization and factorization of chiral Wilson
loops in the AdS description. For the disk we expect three bosonic
zero modes and no fermionic zero modes, while for the annulus, we
expect at least one fermionic zero mode. Notice that for the
straight Wilson loop in figure \ref{figone}(a) we have fermion
zero modes which are the fermionic partners of the bosonic zero
modes for the supersymmetric quantum mechanics along the Wilson
line.  Similar considerations should apply to the closed Wilson loop
correlators discussed in section~\ref{SECONDANSATZ}, which may or may
not have coincident boundaries, but clearly have zero regulated area as
calculated from the $AdS_5 \times S^5$ side of the duality.

\section{Acknowledgements}

We wish to thank A. Kapustin, E. Witten and K. Zarembo for useful
discussions.  Z.~G. thanks the Institute for Advanced Study and the
Boston University visitor program for hospitality during completion
of this work.  The research of A.~D.\ is supported in part by grant
RFBR 04-02-16538, grant for support of scientific schools
NSh-8004.2006.2, and by the National Science Foundation Grant
No.~PHY-0243680.  The research of S.~G.\ is supported in part by the
DOE under Grant No.\ DE-FG02-91ER40671, and by the Sloan Foundation.
The work of J.~M.\ was supported in part by DOE grant
DE-FG01-91ER40654.

\appendix

\section{The Wilson loop operator in the topologically \\ twisted
theory}

In this section we examine the properties of the Wilson loop
operator from the point of view of the topologically twisted
theory. For this purpose it is best to start with ten-dimensional
super Yang Mills and latter reduce it to four dimensions. We
consider the ten-dimensional theory on $\mathbb{C}^5$ and we
consider the supersymmetry associated to the spinor
 that obeys
$\Gamma_{MN} \epsilon = i J_{MN} \epsilon$ in ten dimensions.
Using this spinor we can transform the spinor index on the fermion
into  a vector index, as is usual in complex manifolds. In other
words, we write the gaugino as $\chi = ( \psi + \psi_{ij}
\Gamma^{ij} + \psi_{\bar i} \Gamma^{\bar i} \Gamma^{12345} )
\epsilon$ where $\psi, ~\psi_{ij}, ~\psi_{\bar i}$ are
anticommuting, $\epsilon$ is a commuting spinor and $i,j = 1,
\cdots 5$.

Then the supersymmetry $Q$ associated to $\epsilon$ acts on
the fields as
 \bea
Q \psi &=& g^{i \bar j} {\cal F}_{i \bar j}
 \\
Q \psi_{ij} &=& {\cal F}_{ij}
 \\
 Q \psi_{\bar i} & =&0
 \\
 Q {\cal A}_i & =&0
 \\
 Q  {\cal A}_{\bar i} &=& \psi_{\bar i} \,.
 \eea

We now consider Wilson loop operators which follow a holomorphic
contour
 \be \label{susycont}
 W =\tr P e^{ \int {\cal A}_i dz^i } \,,
 \ee
 where the contour has $\dot z^{\bar i } =0$.
 We see that (\ref{susycont})
 is annihilated by $Q$. On the other hand, if the contour
 $z^i(\tau)$ is topologically trivial, then we see that we can, at
 least formally, expand the operator (\ref{susycont}) in terms of
 local operators which involve ${\cal F}_{ij}$ and holomorphic
 covariant derivatives ${\cal D}_i$.  Such operators are all BRST
 trivial, since ${\cal F}_{ij} = Q \psi_{ij}$ and so are all their
 holomorphic covariant derivatives since ${\cal D}_i$ commutes with $Q$.

In order to go to the four-dimensional ${\cal N}=4$ super Yang
Mills case, all we need to do is to dimensionally reduce by taking
the four-dimensional spacetime to be spanned by the real part of
the first four $z^i$ coordinates. We will then consider contours
with $z^5=0$. The holomorphicity condition $d_\tau z^{\bar i} =0$
translates into the condition (\ref{susy}) relating the tangent
vector to a point on $S^4 \subset S^5$. The complex one-form in
ten dimensions becomes ${\cal A}_i = A_i + i \Phi_i$, in terms of
the four-dimensional gauge field and scalar.

Notice that this argument is very formal. In fact, it fails in the
case of the topological string since in that case the circular
Wilson loop has a nontrivial expectation value. We think that in
four dimensions the result should hold. In fact, we can
continuously deform the 1/16 BPS contour into the 1/4 BPS circular
Wilson loop and then use the arguments in
\cite{GK,Guralnik:2004yc}.

\section{Wilson loops in the topological string }

In this section we examine the question of the Wilson loop in the
topological string context. We study the simplest case which
arises when we consider branes on the deformed conifold and follow
the large $N$ duality to the resolved conifold
\cite{rgcv,hocv,hocvder}

 Let us start with the deformed conifold
 \be
 XY - UV = \epsilon = \sum_{i=1}^4 w_i^2
 \ee
 where we have set $X = w_1 + i w_2$, $Y = w_1 - i w_2$ , $U = w_3+ i
 w_4$ , $V = - w_3 + i w_4$. We have an $S^3$ when all $w_i$ are
 real. We can put three-dimensional branes on this $S^3$.
We can now consider a string worldsheet that lies along the
complex surface $U=V=0$. Such a surface obeys the complex equation
$XY = \epsilon$. This is a noncompact surface which intersects the
$S^3$ at $|Y|= \sqrt{|\epsilon|} $.  We can consider the part of
the original surface that lies along $|Y| \geq \sqrt{|\epsilon|}$.
This is now a worldsheet  that ends on an $S^1 \subset S^3$.
  From the point of view of the
Chern-Simons gauge theory on the $S^3$ this is  a Wilson line in
the fundamental representation. According to the Gopakumar-Vafa
large $N$ transition \cite{rgcv}, this will give rise to a closed
string topological theory on the resolved conifold, which is
specified by the equations
 \be
 \left( \begin{array}{cc} X & U \\ V & Y \end{array} \right)
 \left( \begin{array}{c} \lambda_1 \\ \lambda_2 \end{array} \right) =0
 \ee
where $(\lambda_1, \lambda_2) \in P^1$. There is nonzero solution
only if $XY-UV=0$ and when $X=Y=U=V=0$ we have an additional
solution. We are interested in the surface that is at $U=V=X=0$
and $Y$ non-zero.  This implies that $\lambda_2=0$, so we are at a
point on the $P^1$ as long as $Y\not =0$. This complex surface goes
all the way to the origin. At this point it can wrap or not wrap the
$P^1=S^2$ at the origin.  So we see that we have at least two
choices.\footnote{ In principle we can have multiple wrappings, but
the exact answer shows that they do not contribute.}  One can see
this very explicitly by looking at the    metric for the resolved
conifold \cite{candelas}. In a patch where $\lambda_1 \not =0$, we
can choose complex coordinates $U, Y, \lambda ={\lambda_2 \over
\lambda_1}$. Then the Kahler potential has the form
 \be
 K = f(r^2) + 4 a^2 \log(1 + |\lambda|^2) ~,~~~~~~r^2 \equiv
 (1+|\lambda|^2)(|U|^2 + |Y|^2) \,,
 \ee
 where $a$ is the radius of the two-sphere at the origin.
The simplest choice is for the surface to sit at $U=\lambda =0$.  It is then
spanned by the complex variable $Y$. This obviously obeys the
right boundary conditions and it is a complex surface. In
addition, its area is given by
 \be
 A = \int { 1 \over 2} K_{Y\bar Y} dY d\bar Y = { 1 \over 2}
 F''(r^2) d(r^2) d\varphi = \pi { d F \over d r^2}(r^2_{max}) =
 \pi r_{max}^{4/3} - 2 \pi a^2 \,.
 \ee
 Thus we see that after subtracting an $a$ independent constant
 the regularized area is
 \be
 {\rm Area}_{reg} = - 2 \pi a^2 \,.
 \ee
If we now consider the surface that in addition wraps the $S^2$,
we get $ { \rm Area }_{ reg } =  2 \pi a^2$.  These two
possibilities give rise to the two terms in (\ref{wtop}).


\begin{thebibliography}{99}


\bibitem{Maldacena:1998im}
J.~M.~Maldacena,
Phys.\ Rev.\ Lett.\  {\bf 80}, 4859 (1998) [arXiv:hep-th/9803002].

\bibitem{Rey:1998ik}
S.~J.~Rey and J.~T.~Yee,
Eur.\
Phys.\ J.\ C {\bf 22}, 379 (2001) [arXiv:hep-th/9803001].

\bibitem{Drukker:1999zq}
N.~Drukker, D.~J.~Gross and H.~Ooguri,
Phys.\ Rev.\ D {\bf 60}, 125006 (1999)
[arXiv:hep-th/9904191].

\bibitem{Drukker:2000rr}
N.~Drukker and D.~J.~Gross,
J.\ Math.\ Phys.\  {\bf 42}, 2896
(2001) [arXiv:hep-th/0010274].

\bibitem{morecomplex}
  N.~Drukker and B.~Fiol,
  JHEP {\bf 0502}, 010 (2005)
  [arXiv:hep-th/0501109].
S.~Yamaguchi,
  arXiv:hep-th/0601089;
    hep-th/0603208.
  J.~Gomis and F.~Passerini,
  arXiv:hep-th/0604007.


\bibitem{Zarembo:2002an}
K.~Zarembo,
Nucl.\ Phys.\ B {\bf 643}, 157 (2002) [arXiv:hep-th/0205160].

\bibitem{vafageom}
  C.~Vafa,
  J.\ Math.\ Phys.\  {\bf 42}, 2798 (2001)
  [arXiv:hep-th/0008142].
\bibitem{GK}
  Z.~Guralnik and B.~Kulik,
  JHEP {\bf 0401}, 065 (2004)
  [arXiv:hep-th/0309118].

\bibitem{Guralnik:2004yc}
  Z.~Guralnik, S.~Kovacs and B.~Kulik,
  arXiv:hep-th/0409091.

\bibitem{Joyce:2001xt}
  D.~Joyce,
  arXiv:math.dg/0108088.

\bibitem{Lu:1998nu}
H.~Lu, C.~N.~Pope and J.~Rahmfeld,
J.\ Math.\ Phys.\  {\bf 40}, 4518 (1999)
[arXiv:hep-th/9805151].

\bibitem{sofourtwist}
It is case $i)$ in
 C.~Vafa and E.~Witten,
  Nucl.\ Phys.\ B {\bf 431}, 3 (1994)
  [arXiv:hep-th/9408074],  with a correction mentioned in page 16
  of \cite{vafatwist}

\bibitem{vafatwist}
  M.~Bershadsky, C.~Vafa and V.~Sadov,
  Nucl.\ Phys.\ B {\bf 463}, 420 (1996)
  [arXiv:hep-th/9511222].

\bibitem{amv}
  M.~Atiyah, J.~M.~Maldacena and C.~Vafa,
  J.\ Math.\ Phys.\  {\bf 42}, 3209 (2001)
  [arXiv:hep-th/0011256].

\bibitem{candelas}
  P.~Candelas and X.~C.~de la Ossa,
  Nucl.\ Phys.\ B {\bf 342}, 246 (1990).

\bibitem{hocv}
  H.~Ooguri and C.~Vafa,
  Nucl.\ Phys.\ B {\bf 577}, 419 (2000)
  [arXiv:hep-th/9912123].

\bibitem{hocvder}
  H.~Ooguri and C.~Vafa,
  Nucl.\ Phys.\ B {\bf 641} (2002) 3
  [arXiv:hep-th/0205297].


\bibitem{rgcv}
  R.~Gopakumar and C.~Vafa,
  Adv.\ Theor.\ Math.\ Phys.\  {\bf 3}, 1415 (1999)
  [arXiv:hep-th/9811131].

\bibitem{wittenjones}
  E.~Witten,
  Commun.\ Math.\ Phys.\  {\bf 121}, 351 (1989).

\bibitem{BrezisCoron}
  H.~Brezis and J.~M.~Coron,
  Comm.~Math.~Phys {\bf 92} (1983), 203-215.

\bibitem{BenciCoron}
  V. Benci and J.~M.~Coron,
  Ann.~Inst. Henri Poincaré, Anal. Non Linéaire 2, 119-141 (1985). [ISSN 0294-1449]

\bibitem{Brezis}
  H.~Brezis,
  Bull.~Am.~Math.~Soc {\bf 40} no.2 (2003), 179-201.

\bibitem{DrukkerFiol}
  N.~Drukker and B.~Fiol,
  JHEP {\bf 0601}, 056 (2006)
  [arXiv:hep-th/0506058].

\bibitem{russo}
  G.~Arutyunov, S.~Frolov, J.~Russo and A.~A.~Tseytlin,
  Nucl.\ Phys.\ B {\bf 671}, 3 (2003)
  [arXiv:hep-th/0307191].

\bibitem{ancv}
  A.~Neitzke and C.~Vafa,
  arXiv:hep-th/0410178.

\bibitem{uranga}
  F.~Canoura, J.~D.~Edelstein, L.~A.~P.~Zayas, A.~V.~Ramallo and D.~Vaman,
  arXiv:hep-th/0512087.
  J.~F.~G.~Cascales and A.~M.~Uranga,
  JHEP {\bf 0411}, 083 (2004)
  [arXiv:hep-th/0407132].


\bibitem{gaunt}
  B.~S.~Acharya, J.~P.~Gauntlett and N.~Kim,
  Phys.\ Rev.\ D {\bf 63}, 106003 (2001)
  [arXiv:hep-th/0011190].

\bibitem{JMCN}
  J.~M.~Maldacena and C.~Nunez,
  Int.\ J.\ Mod.\ Phys.\ A {\bf 16}, 822 (2001)
  [arXiv:hep-th/0007018].



\end{thebibliography}
\end{document}